\documentclass[twocolumn,pra,showpacs,amsmath,amssymb]{revtex4}

\usepackage{amssymb,amsmath,amsfonts,amsthm,graphicx,color}

\begin{document}

\title{Bounding the dimension of bipartite quantum systems}
\author{Tam\'as V\'ertesi}
\email{tvertesi@dtp.atomki.hu}
\author{K\'aroly F.~P\'al}
\email{kfpal@atomki.hu}
\affiliation{Institute of Nuclear Research of the Hungarian Academy of Sciences\\
H-4001 Debrecen, P.O.~Box 51, Hungary}

\def\CC{\mathbb{C}}
\def\RR{\mathbb{R}}
\def\one{\leavevmode\hbox{\small1\normalsize\kern-.33em1}}
\newcommand*{\tr}{\mathsf{Tr}}
\newcommand{\diag}{\mathop{\mathrm{diag}}}
\newcommand{\ket}[1]{|#1\rangle}
\newcommand{\bra}[1]{\langle#1|}
\newtheorem{theorem}{Theorem}
\newtheorem{lemma}{Lemma}

\date{\today}

\begin{abstract}
Let us consider the set of joint quantum correlations arising from
two-outcome local measurements on a bipartite quantum system. We
prove that no finite dimension is sufficient to generate all these
sets. We approach the problem in two different ways by
constructing explicit examples for every dimension $d$, which
demonstrates that there exist bipartite correlations that
necessitate $d$-dimensional local quantum systems in order to
generate them. We also show that at least 10 two-outcome
measurements must be carried out by the two parties altogether so
as to generate bipartite joint correlations not achievable by
two-dimensional local systems. The smallest explicit example we
found involves 11 settings.
\end{abstract}

 \maketitle

\section{Introduction}\label{intro}

We consider the following standard bipartite measurement scenario
\cite{Bell64}. Two distant, separated parties, conventionally
called Alice and Bob, share a joint quantum state $\rho$. The two
participants carry out measurements on their respective states and
may obtain two possible classical outcomes $\pm 1$. Let us specify
the situation, that Alice (Bob) has $m_A$ ($m_B$) number of
$\pm$1-valued observables $A_1,\ldots,A_{m_A}$
($B_1,\ldots,B_{m_B}$). In a particular run of the experiment
Alice and Bob each measures one observable, $A_i$ and $B_j$,
getting the respective outputs $a_i\in\pm 1$ and $b_j\in\pm 1$ and
then makes the product $a_i b_j$. By repeating this process many
times they can form the joint correlation
\begin{equation}
E(a_i,b_j)=\langle A_i B_j\rangle = \tr(\rho A_i\otimes B_j), \label{joint}
\end{equation}
which is the expected value of the product $a_i b_j$. These joint
correlations for all $\{i,j\}$ form an $m_A m_B$ dimensional
vector, designated by $\mathbf{c}$.

We are interested in the following problem. We are given a vector
$\mathbf{c}$ of joint correlations $E(a_i,b_j)$, with
$i=1,\ldots,m_A$ and $j=1,\ldots,m_B$, for a particular number of
settings $m_A$ and $m_B$. What is the minimum Hilbert-space
dimension which is needed to reproduce this vector $\mathbf{c}$?
Is it the case that all sets of these vectors for any $m_A$ and
$m_B$ are possibly reproduced by two-dimensional local Hilbert
spaces? In a recent work we provided counterexamples for small
values of $m_A$ and $m_B$ (numerically for $m_B=4$ and $m_A=8$; an
exact treatment was carried out for $m_B=6$ and $m_A=15$), which
shows that indeed higher than two-dimensional local Hilbert spaces
are required in order to produce all the set of such vectors
\cite{VP08}. The same conclusion, based on the gap between a lower
bound on the Grothendieck constant in infinite dimensions
\cite{FR94} and an upper bound on the three-dimensional
Grothendieck constant \cite{Krivine79}, has been obtained by
Brunner et al. \cite{Brunner}. In this paper the idea of dimension
witnesses has been put forward as well. Brunner et al.
\cite{Brunner} defined a $d$-dimensional witness as a certain kind
of generalization of Bell inequalities \cite{Bell64}, such that
all quantum correlations arising from observables on
$d$-dimensional component Hilbert spaces satisfy the inequality.
Hence, correlations which violate the inequality could be
established only by measuring systems of dimension larger than
$d$. Thus dimension witness is a useful tool for measuring the
dimension of multipartite quantum systems relying only on
experimental data.

The results in \cite{Brunner} and \cite{VP08} prove the existence
of two-dimensional witnesses in bipartite systems based on only
joint correlations. On the other hand, Perez-Garcia et
al.~\cite{Perez} gave a non-constructive proof of the existence of
$d$-dimensional witnesses for tripartite systems for any dimension
$d$. In particular, they proved in Ref.~\cite{Perez} in their
Theorem~1 that for every dimension $d$, there exists a dimension
$D$, a pure state $|\psi\rangle$ on $\CC^d\otimes \CC^d \otimes
\CC^D$, and a Bell inequality with $\pm 1$-valued observables such
that the violation by $|\psi\rangle$ is greater than $\sqrt d$ up
to some universal constant. We also mention a recent paper by
Wehner et al. \cite{WCD}, in which a method has been given to
bound from below the dimension of a Hilbert space by relating this
problem to the construction of quantum random access codes. Their
method works for quantum systems involving any number of parties
(can be applied even for a single system) and is most useful for
small number of settings.

In this paper we tackle the problem of whether there exist
dimension witnesses for bipartite systems built from two-outcome
measurements for any dimension $d$. Note, that the tripartite case
has been affirmatively answered in \cite{Perez}. On the other
hand, the fact that this may also hold for the bipartite case has
been conjectured in Ref.~\cite{Brunner} based on the plausible
assumption that the Grothendieck constant \cite{Finch} for
infinite dimension is strictly larger than the Grothendieck
constant for every finite order. In the numerical works of
\cite{PV08a,PV08b} we found various dimension witnesses up to
dimensions of three and four, respectively, using bipartite Bell
inequalities up to five binary measurement settings $m_A=m_B=5$.
But there, besides joint correlations, local marginal terms were
also involved.

In particular, in Sec.~\ref{rep} of the present work, we discuss
the link between joint correlations of two-party outcomes arising
from two-outcome measurements and dot products of unit vectors in
the Euclidean space. Armed with this connection two families of
Bell inequalities are presented (in Sec.~\ref{finite} and in
Sec.~\ref{cont}), proving the existence of bipartite correlations
for any dimension $d$ which require at least $d$-dimensional
systems in order to generate them. The first family
(Sec.~\ref{finite}) involves a finite number of settings for any
finite $d$, while in case of the other family the number of
settings is infinite. In Sec.~\ref{finite}, as a by-product, it is
shown analytically that joint correlations of two parties
resulting from two-outcome measurements, which require more than
two-dimensional component spaces, cannot be generated by $m_B=4$
and $m_A=6$ or by less settings. That is, these numbers of
settings are not enough to define a two-dimensional witness. On
the contrary, examples are given for a two-dimensional witness for
settings $m_B=4$ and $m_A=7$ and for settings $m_B=5$ and $m_A=6$.
The paper concludes with Sec.~\ref{conc}.

\section{Representation of joint correlations with dot products}\label{rep}

This section contains two lemmas. The first one is borrowed from
\cite{AGT06} (their Lemma~2), establishing a link between dot
product of unit vectors in the Euclidean space and joint
correlations originating from projective measurements in the
Hilbert space, whereas the second one is a strengthening of this
lemma regarding dimension witnesses.

Here we state the first lemma without proof.

\begin{lemma}(Ac\'\i n et al. \cite{AGT06})
Suppose Alice and Bob measure observables $A$ and $B$ on a pure
quantum state $\ket \psi \in \CC^d \otimes \CC^d$. Then we can
associate a real unit vector $\vec a\in \RR^{2d^2}$ with $A$
(independent of ${B}$), and a real unit vector $\vec b \in
\RR^{2d^2}$ with $B$ (independent of $A$) such that $ E(a,b)= \vec
a \cdot \vec b$. Moreover, if $\ket \psi$ is maximally entangled,
then we can assume that the vectors $\vec a$ and $\vec b$ lie in
$\RR^{d^2-1}$. \label{lemmaAGT}
\end{lemma}

Let us fix some notations regarding dimension witnesses
\cite{Brunner} based on joint correlations. A $d$-dimensional
witness \cite{Brunner}, in terms of joint correlations
(\ref{joint}), is a linear function of these correlators,
described by a vector $\mathbf{m}$ of real coefficients $M_{ij}$,
such that
\begin{equation}
\mathbf{m}\cdot \mathbf{c}=\sum_{i=1}^{m_A}\sum_{j=1}^{m_B}M_{ij}E(a_i,b_j)\le W^d, \label{dimformula}
\end{equation}
for all correlations of form (\ref{joint}) with $\rho$ in
$\CC^d\otimes \CC^d$, and such that there are joint correlations
for higher than $d$-dimensional systems for which $\mathbf{m}\cdot
\mathbf{c}>W^d$. Note, that due to convexity arguments it is
sufficient to consider only pure states
$\rho=|\psi\rangle\langle\psi|$ in order to achieve the maximum on
the LHS of (\ref{dimformula}).

Lemma~\ref{lemmaAGT} implies the following. Let us take an
arbitrary vector $\mathbf{m}$ in (\ref{dimformula}) defined by
some coefficients $M_{ij}$, and denote the maximum value
\begin{equation}
Q^d=\max_{\mathbf{c}}{\{\mathbf{m}\cdot \mathbf{c}\}}
\label{defQd}
\end{equation}
achievable by correlations~(\ref{joint}) from $\CC^d\otimes
\CC^d$. Because of Lemma~\ref{lemmaAGT} we can associate with the
elements of $\mathbf{c}$, $E(a_i,b_j)$, two normalized vectors
$\vec a_i, \vec b_j \in \RR^n$ such that $E(a_i,b_j)=\vec a_i
\cdot \vec b_j$. Defining
\begin{equation}
{\cal B}=\sum_{i=1}^{m_A}\sum_{j=1}^{m_B}M_{ij}\vec a_i\cdot\vec b_j, \label{corr}
\end{equation}
and denoting the extremal value in the $n$-dimensional Euclidean
space by
\begin{equation}
T^n = \max_{\vec a_i, \vec b_j\in \RR^n}\mathcal{B}\label{maxcorr}. \end{equation}
Lemma~\ref{lemmaAGT} implies that $Q^d\le T^n$ for $n=2d^2$.

This way we can bound the quantum value $Q^d$ achievable in
$d$-dimensional component Hilbert spaces from above by dot product
of unit vectors in $\RR^n$. On the other hand, due to Tsirelson's
construction \cite{Tsirelson}, all dot products of unit vectors,
$\vec a, \vec b \in \RR^m$, can be realized as $\pm 1$-valued
observables on a maximally entangled state on $\CC^D\otimes\CC^D$,
where $D=2^{\lfloor m/2\rfloor}$. Then in light of this and the
fact that $Q^d\le T^n$ for $n=2d^2$, constructing a Bell
expression~(\ref{corr}) for which $T^n<T^{m}$, with $n<m$, implies
a $d$-dimensional witness. The next Lemma sharpens this statement.

\begin{lemma}
In the notation defined above by (\ref{defQd}) and (\ref{mc}), we
have $Q^d\le T^n$ for $n=2d-1$. Furthermore, if the Hilbert space
is restricted to be real, then the above relation holds true with
$n=d$. Thus, by constructing a Bell expression $\mathcal{B}$ for
which $T^n<T^{n+1}$ implies a $d$-dimensional witness with
$d=\lfloor\frac{n+1}{2}\rfloor$. \label{lemmasharp}
\end{lemma}

In Sec.~\ref{finite} we present a family of Bell expressions,
characterized by Bob's number of settings $n=m_B$, for which one
can prove, that the above condition $T^n<T^{n+1}$ holds for the
($n+1$)th member of the family. In case of this family, however,
we can give the gap $(T^{n+1}-T^n)$ explicitly only for small $n$
values. In Sec.~\ref{cont} another Bell expression is provided,
where this gap can be calculated analytically for every $n$.

\begin{proof}
Here we prove Lemma~\ref{lemmasharp} and start by proving the
relation $Q^d\le T^d$ in the case of Hilbert space $\CC^d\otimes
\CC^d$ restricted to be real. Since $A^i$ and $B^j$ are $\pm
1$-valued observables the normalization conditions
$\sum_{l=1}^d{(a_{kl}^i)^2}=1$ and $\sum_{l=1}^d{(b_{kl}^j)^2}=1$
are satisfied for all $k=1,\ldots,d$, where $A^i=(a_{kl}^i)$ and
$B^j=(b_{kl}^j)$ are $d\times d$ symmetric matrices with real
coefficients (designations $i$ and $j$ are moved to upper indices
for convenience). On the other hand, for given $\mathbf{m}$,
$Q^d=\max_{\mathbf{c}}{\{\mathbf{m}\cdot\mathbf
{c}\}}=\max{\sum_{ij}{M_{ij}\langle \psi|A^i\otimes
B^j|\psi\rangle}}$, where maximization is over all
$|\psi\rangle\langle\psi|$ in $\CC^d\otimes\CC^d$ and for all
observables $A^i$ and $B^j$ in $\CC^d$. The state $|\psi\rangle$
can be written in Schmidt form,
$|\psi\rangle=\sum_{i=1}^d{\alpha_i}|ii\rangle$, where $\alpha_i$
can be chosen positive and their squares add up to unity. Then we
have
\begin{equation}
\mathbf{m}\cdot\mathbf{c}=\sum_{ij}{M_{ij}\langle \psi|A^i\otimes
B^j|\psi\rangle}=\sum_{kl}{\alpha_k \alpha_l N_{kl}}, \label{mc}
\end{equation}
where $N_{kl}=\sum_{ij}{M_{ij}(a_{kl}^i b_{kl}^j + a_{lk}^i
b_{lk}^j})/2$. Note that in the case of real matrices $A^i$ and
$B^j$ the products $a_{kl}^i b_{kl}^j$ and $a_{lk}^i b_{lk}^j$ are
the same. However, this form will turn out to be useful in the
proof of the complex case. Now let us flip the signs of $a_{kl}^i$
and $a_{lk}^i$ for all $i=1,\ldots,d$ in each pair $\{k,l\}$ if
$N_{kl}$ is negative. By performing the necessary sign flips we
obtain the matrices $\tilde A^i$ defined by elements $\tilde
a_{kl}^i$. In this way the sum $\tilde N_{kl}=\sum_{ij}{M_{ij}
(\tilde a_{kl}^i b_{kl}^j + \tilde a_{lk}^i b_{lk}^j)/2}$ becomes
positive for any given $\{k,l\}$ pair. Thus, as a result the value
of $\mathbf{m}\cdot\mathbf{c}$ cannot decrease. Furthermore the
normalization condition still holds for the sign-flipped matrices,
$\sum_{l=1}^d{(\tilde a_{kl}^i)^2}=1$. In the following, through a
chain of inequalities we obtain an upper bound for $Q^d$. First of
all, we can write
\begin{align}
\sum_{kl}{\alpha_k\alpha_l N_{kl}}&\le \sum_{kl}{(\alpha_k^2+\alpha_l^2)\tilde N_{kl}}/2\label{chain1}\\
&= \sum_{kl}{\alpha_k^2 \sum_{ij}{M_{ij} (\tilde a_{kl}^i
b_{kl}^j + \tilde a_{lk}^i b_{lk}^j)}}/2\label{chain2}\\
&=\sum_k{\alpha_k^2 \sum_{ij}{M_{ij}\vec a_k^i\cdot \vec
b_k^j}}\label{chain3}\\&\le \sum_k{\alpha_k^2\sum_{ij}M_{ij}{\vec
a^i \cdot\vec b^j}}= \sum_{ij}M_{ij}{\vec a^i \cdot\vec b^j},
\label{chain}
\end{align}
where the unit vectors $\vec a_k^i, \vec b_k^j$ and $\vec a^i,
\vec b^j$ are in the $d$-dimensional Euclidean space. In
inequality~(\ref{chain1}) we used $N_{kl}\le \tilde N_{kl}$ and
the relation $2xy\le x^2+y^2$ holding for any real number $x$ and
$y$. Equality~(\ref{chain2}) follows from the fact that the
matrices $\tilde A^i$ and $B^j$ are symmetric. In
equality~(\ref{chain3}) we exploited that $\tilde a_{kl}^i$ and
$b_{kl}^j$ can be treated as the $d$ real components of unit
vectors $\vec a_k^i=(\tilde a_{k1}^i, \tilde a_{k2}^i,\ldots,
\tilde a_{kd}^i)$ and $\vec b_k^j=(b_{k1}^j, b_{k2}^j,\ldots,
b_{kd}^j)$. In the last inequality we omitted index $k$, but
keeping in mind the normalization conditions for $\vec a_k^i$ and
$\vec b_k^j$ we have $\vec a^i\cdot \vec a^i =1$ and $\vec
b^j\cdot \vec b^j =1$. Finally the last equality owes to
$\sum_k{\alpha_k}^2=1$. According to (\ref{defQd}) and (\ref{mc}),
$Q^d$ is the maximum of $\sum_{kl}{\alpha_k\alpha_l N_{kl}}$ over
all observables and states of dimension $d$. Thus, by use of the
chain of inequalities~(\ref{chain}), we obtain the upper bound
\begin{equation}
Q^d\le \max_{\vec a_i,\vec b_j\in R^{d}}{\sum_{ij}M_{ij}{\vec a^i
\cdot\vec b^j}}=T^d, \label{Qd}
\end{equation}
where $\vec a_i$ and $\vec b_j$ are unit vectors.

The proof regarding the complex case goes along the same line as
in the real case, but in this case $A^i=(a_{kl}^i)$ and
$B^j=(b_{kl}^{*j})$ are by definition Hermitian matrices and their
components are complex valued, where $*$ denotes complex
conjugation. Similarly as in the real case, $\tilde a_{kl}^i$ is
defined by flipping the sign of $a_{kl}^i$ if $N_{kl}$ is negative
for a given pair $(k,l)$. Then in the complex case, in place of
the real number $(\tilde a_{kl}^i b_{kl}^j + \tilde a_{lk}^i
b_{lk}^j)$ one can write $2(\mathrm{Re}\tilde a_{kl}^i \mathrm{Re}
(b_{kl}^{j}) - \mathrm{Im}\tilde a_{kl}^i \mathrm{Im}
(b_{kl}^{j}))$, and then in (\ref{chain3}) we have $\vec a_k^i =
(\mathrm{Re} \tilde a_{k1}^i,\mathrm{Im} \tilde
a_{k1}^i,\ldots,\tilde a_{kk}^i,\ldots,\mathrm{Re} \tilde
a_{kd}^i,\mathrm{Im} \tilde a_{kd}^i)$ and $\vec b_k^j =
(\mathrm{Re} b_{k1}^j,-\mathrm{Im} b_{k1}^j,\ldots,
b_{kk}^j,\ldots,\mathrm{Re} \tilde b_{kd}^j,-\mathrm{Im} \tilde
b_{kd}^j)$. Since components $\tilde a_{kk}^i$ and $b_{kk}^j$ are
real, these vectors lie in the $(2d-1)$-dimensional Euclidean
space, and they can be checked to be of unit length. Therefore, in
the complex case in ~(\ref{Qd}) one has the unit vectors $\vec
a^i, \vec b^j\in \RR^{2d-1}$ and the RHS becomes $T^{2d-1}$, which
completes the proof.
\end{proof}

Some remarks about Lemma~\ref{lemmasharp} are in order. First,
provided that the real Hilbert-space result is given, one can also
obtain the result concerning the complex case by mapping the
$d\times d$ Hermitian observables to $2d\times 2d$ real
observables (as discussed in the Appendix of \cite{PV08a} and in
the multipartite setting in \cite{MMG}), which latter matrices
have the property that at each column there is at least one zero
component, entailing the $(2d-1)$-dimensional Euclidean vectors
$\vec a^i$ and $\vec b^j$ in the upper bound expression on $Q^d$.
Second, in light of the proofs in Refs.~\cite{Brunner,VP08}
regarding the qubit case, in Lemma~\ref{lemmasharp}, $Q^2$ is
equal to $T^3$ and $T^2$ in the respective cases of complex and
real qubits.

\section{Bounds on dimensions with finite number of settings}\label{finite}

In this section an example for dimension witness is provided for
every dimension $d$. Let us consider the expression $\cal B$
defined earlier in Sec.~\ref{rep} by (\ref{corr}). According to
Lemma~\ref{lemmasharp}, a Bell expression with a specific matrix
$\mathbf{m}$ for which $T^n<T^{n+1}$ ($T^n$ denoting the maximum
value of (\ref{corr}) in $\RR^n$) implies the existence of a
$d$-dimensional Hilbert-space witness in a bipartite system with
$d=\lfloor\frac{n+1}{2}\rfloor$.

In Sec.~\ref{limits} it is shown that if $m_A\le m_B(m_B-1)/2$,
then vectors from a smaller space than dimension $m_B$ are enough
to maximize $\cal B$. Then in Sec.~\ref{bell} it is shown via a
particular family of Bell inequalities (labeled by Bob's
measurements $m_B$) that when $m_A=m_B(m_B-1)/2+1$ for any given
value of $m_B$, the whole $m_B$-dimensional Euclidean space is
required to maximize $\cal B$. This proves for every $n=m_B-1$
that $T^n<T^{n+1}$, and implies that the Bell coefficients
$\mathbf{m} = (M_{ij})$ of this particular family define a
dimension witness in Hilbert-space dimension
$d=\lfloor\frac{m_B}{2}\rfloor$, where $m_B$ is the number of
settings on Bob's side. In Sec.~\ref{twodim} we determine the gap
between the maximum value achievable in $m_B$ and in
$(m_B-1)$-dimensional Euclidean spaces for small $m_B$, and we
also discuss the question of minimal number of settings in order
for witnesses for two-dimensional component spaces to exist.

\subsection{Limits on the number of settings}\label{limits}

Let us start from expression~(\ref{corr}). Without the loss of
generality, we may suppose that $m_B\le m_A$. The dimensions of
the real vector spaces are large enough to accommodate vectors
that maximize the expression. From
\begin{equation}
{\cal B}=\sum_{i=1}^{m_A}\vec a_i\cdot\left(\sum_{j=1}^{m_B}M_{ij}\vec b_j\right)
\end{equation}
it is clear that when ${\cal B}$ is maximal, each $\vec a_i$
points to the same direction as the linear combination of the
$\vec b_j$ vectors it is multiplied with. Therefore, however large
the number of Alice's  measurement settings $m_A$ is, her vector
space need not have a larger dimensionality than that of Bob.
Bob's $m_B$ vectors can always be accommodated in an
$m_B$-dimensional space. What we will show here, is that if
$m_A<m_B(m_B-1)/2+1$, then even vectors from a smaller space are
enough to maximize $\cal B$.

Let $\bar{\cal B}$ be the same as $\cal B$ with Alice's vectors
$\vec a_i$ chosen optimally. $\bar{\cal B}$ is only the function
of Bob's vectors $\vec b_j$:
\begin{equation}
\bar{\cal B}=\sum_{i=1}^{m_A}\left|\sum_{j=1}^{m_B}M_{ij}\vec b_j\right| \equiv\sum_{i=1}^{m_A}l_i.
\end{equation}
We can always find the maximum of $\cal B$ by maximizing
$\bar{\cal B}$ in terms of the unit vectors $\vec b$ of an
$m_B$-dimensional vector space. The terms of the above equation
can further be written as
\begin{equation}
l_k=\sqrt{\sum_{i}M_{ki}^2+2\sum_{i>j}M_{ki}M_{kj}X_{ij}},
\end{equation}
where
\begin{equation}
X_{ij}\equiv\vec b_{i} \cdot \vec b_{j}.
\end{equation}
The $X_{ij}$, $(j<i\le m_B)$ values determine the relative
directions of the unit vectors and, therefore, up to an irrelevant
orthogonal transformation, the vectors themselves. We will regard
$\bar{\cal B}$ as a function of these $N\equiv m_B(m_B-1)/2$
numbers. Besides $|X_{ij}|\le 1$, the variables must satisfy
several other constraints. The domain of the function is where the
Gramian matrix, the symmetric matrix containing $X_{ij}$ and
$X_{ii}=1$ is positive semidefinite. Its determinant, the Gramian
is zero at the boundary of the domain, which means that the
vectors $\vec b_j$ are linearly dependent. If the $X_{ij}$
variables maximizing $\bar{\cal B}$ lie there, then a space of
less than $m_B$ dimensions is enough to accommodate the
corresponding vectors. The whole $m_B$-dimensional space is only
required if the maximum is not at the boundary of the domain. The
$\bar{\cal B}$ as a function of $X_{ij}$ is non-analytical only
where $l_i=0$, but that occurs only at the boundary, as $l_i=0$
means linear dependence of the vectors.

From these it follows that if the vectors $\vec b_j$ maximizing
$\bar{\cal B}$ span the $m_B$-dimensional space, then all partial
derivatives of the function must vanish there. Let us introduce a
single index denoted by a Greek letter instead of the pair
$\{i,j\}$, say let there be $\nu=(i-1)(i-2)/2+j$. We also
introduce the following notation: $x_{\nu}\equiv X_{ij}$,
$Y_{k\nu}\equiv M_{ki}M_{kj}$, and $C_k\equiv \sum_{i}M_{ki}^2$.
With this notation
\begin{equation}
\label{lksimp} l_k=\sqrt{C_k+2\sum_{\nu=1}^{N}Y_{k\nu}x_{\nu}}.
\end{equation}
At the maximum of $\bar{\cal B}$
\begin{equation}
\frac{\partial\bar{\cal B}}{\partial x_\alpha}=\frac{\partial}{\partial x_\alpha}
\sum_{k=1}^{m_A}l_k=\sum_{k=1}^{m_A}Y_{k\alpha}\frac 1{l_k}=0
\end{equation}
must hold. This is a set of $N$ homogenous linear equations for
$1/l_k$. If the number of variables $m_A$  is less than $N+1$, the
equations may only have nontrivial solution if no more than
$m_A-1$ of them are independent. If such a solution exists, it may
define $l_k$ only up to a constant factor. Therefore,
$l_k=y/\lambda_k$, where $\lambda_k$ is a nontrivial solution. All
$\lambda_k$ must be nonzero and must have the same sign, because
$l_k$ must be positive. If these conditions hold, from
Eq.~(\ref{lksimp}) we get a set of $m_A$ linear equations for
$y^2$ and $x_\nu$:
\begin{equation}
2\sum_{\nu=1}^{N}Y_{k\nu}x_{\nu}-y^2/\lambda_k^2=-C_k.
\end{equation}
The solution of these equations can only correspond to a solution
of our problem, if $y^2$ is positive, $x_{\nu}$ are within the
boundary of the domain and if it defines a maximum. However, if
all these are true, the solution is not unique if the number of
equations $m_A$ is less than the number of variables
$N+1=m_B(m_B-1)/2+1$. In this case the equations are satisfied on
a whole subspace of the space containing the domain of $\bar{\cal
B}$. The function is constant in that subspace up to the boundary
of the domain. Therefore, where the subspace crosses the boundary,
the function will still have its maximum value. This way
$\bar{\cal B}$ can be maximized in a less than $m_B$-dimensional
space.

To summarize the argument, if $m_A$ is not large enough, the
equations requiring the partial derivatives to be zero usually
have no solution corresponding to a maximum. In this case the
function will take its maximum value with its variables at the
boundary of their domain. With the Bell factors $M_{ij}$ having
some very special values the equations may be solvable, but then
they cannot be independent. Therefore they cannot define the
variables in a unique way. The function will have the same value
on a subset of its domain, extending to the boundary. This means
that either there is no solution or there is an infinity of them.
Only vectors spanning a less than $m_B$-dimensional space will
maximize $\bar{\cal B}$ in the first case, and there will exist
such vectors maximizing $\bar{\cal B}$ in the second case.

\subsection{Family of Bell expressions}\label{bell}

When $m_A=m_B(m_B-1)/2+1=N+1$ then the whole $m_B$-dimensional space may be required to maximize $\bar{\cal B}$. An example of such a case is the following:
\begin{equation}
\mathcal{B}\equiv \sum_{m_B\ge i>j\ge 1}\vec a_{ij}(\vec b_i-\vec b_j) + \gamma \vec
a_0\sum_{i=1}^{m_B}\vec b_i, \label{Bgamma}
\end{equation}
where the corresponding $M_{ij}$ is defined through
Eq.~(\ref{corr}), parametrized by $m_B$, and $\gamma>0$. Note that
setting $\gamma=0$, we obtain the $Z_n$ family introduced in
\cite{VP08}. By choosing Alice's vectors optimally,
\begin{align}
\bar{\cal B}&=\sum_{i>j}\left|\vec b_i-\vec b_j\right|+\gamma\left|\sum_{i=1}^{m_B} \vec b_i\right|\nonumber\\
&=\sum_{\mu=1}^N\sqrt{2-2x_\mu}+\gamma\sqrt{m_B+2\sum_{\nu=1}^Nx_\nu} \label{eq:koczka}
\end{align}
Here we used the single Greek letter notation for the $\{i,j\}$ pair we introduced earlier. The
partial derivatives of equation (\ref{eq:koczka}) are:
\begin{equation}
\frac{\partial\bar{\cal B}}{\partial x_\alpha}=-\frac{1}{\sqrt{2-2x_{\alpha}}}+
\frac{\gamma}{\sqrt{m_B+2\sum_{\nu=1}^Nx_\nu}},
\end{equation}
which are zero at
\begin{equation}
x_\alpha=x=\frac{2\gamma^2-m_B}{2\gamma^2+m_B(m_B-1)}.
\end{equation}
Thus at the extremal point all nondiagonal entries of the Gramian
will be equal, $x<1$, and the diagonal ones are all 1 (the $\vec
b_i$ are normalized). Let us consider a determinant of size
$k\times k$ with all diagonal elements having the value of $p$ and
all nondiagonal ones are equal to $q$. Let us subtract the
$(i-1)$th column from the $i$th one in turn for
$i=k,k-1,\dots,3,2$, in this order. Then each column, except for
the unchanged first one, will contain $(p-q)$ in the diagonal,
$(q-p)$ just above it, and zero everywhere else. Then add the
$i$th row to the $(i-1)$th in turn for $i=k,k-1,\dots,3,2$, in
this order. This way only the diagonal element will be nonzero in
each column but the first one, with a value of $(p-q)$, while the
first element of the first column will be the sum of all original
entries in that column, that is, $p+(k-1)q$. Therefore, the value
of the determinant, which is now the product of all diagonal
entries, will simply be $[p+(k-1)q](p-q)^{k-1}$. With $p=1$ and
$q=x>-1/(m_B-1)$, which is true if $\gamma>0$, the determinant is
positive for any $k\le m_B$. Therefore, the Gramian is positive
definite, which means that we are not at the boundary of the
domain. This is the only extremal point, and at that point
\begin{equation}
\bar{\cal B}=T^{m_B}=m_B\sqrt{\gamma^2+m_B(m_B-1)/2}. \label{Bmax}
\end{equation}
The second derivatives of equation (\ref{eq:koczka}) are
\begin{equation}
\frac{\partial\bar{\cal B}}{\partial x_\alpha\partial x_\beta}=
-(2-2x_{\alpha})^{-3/2}\delta_{\alpha\beta}-\gamma\left(m_B+2\sum_{\nu=1}^Nx_\nu\right)^{-3/2}
\end{equation}
At the extremal point all nondiagonal elements $\alpha\neq\beta$
of $-1$ times the matrix are equal,
$q'=\gamma[m_B+m_B(m_B-1)x]^{-3/2}>0$, and also all diagonal
elements $\alpha=\beta$ are equal, $p'=q'+(2-2x)^{-3/2}> q'$. From
the values of the determinants of matrices with such entries given
above, it follows that minus one times the second derivative matrix
is positive definite at the extremum; therefore the extremum is a
true maximum. There is no other extremal point in the domain of
the function, so this one has to be the absolute maximum of
$\bar{\cal B}$. This proves $T^n<T^{n+1}$ for Bell
expression~(\ref{Bgamma}) for $n=m_B-1$, entailing a dimension
witness $d=\lfloor m_B/2\rfloor$ for every $m_B$ according to
Lemma~\ref{lemmasharp}.

The $\gamma=\sqrt{m_B/2}$ is an interesting special value, in
which case $x=0$; that is, all $\vec b_i$ vectors are orthogonal to
each other. Geometrically, we may consider $\vec b_i$ as edges, and
$\vec b_i-\vec b_j$ as face diagonals of a hypercube, while
$\sum_{i=1}^{m_B}\vec b_i$ is a vector pointing toward one of its
space diagonals. If we decrease $\gamma$, that is, the weight of
this space diagonal in the expression to be maximized, the
geometrical object will flatten along this direction, and at
$\gamma=0$ it will collapse to become an $(m_B-1)$-dimensional
object. The corresponding Bell inequality has one less measurement
settings for Alice, and it belongs to family $Z$ in
Ref.~\cite{VP08}. When $\gamma\neq 0$, the one extra term is
enough to prevent the collapse, and the whole $m_B$-dimensional
space is required to accommodate the optimum object.

We can determine the classical limit as the maximum value of
\begin{equation}
\bar{\cal B}=\sum_{i>j}\left|z_i-z_j\right|+\gamma\left|\sum_{i=1}^{m_B}z_i\right|,
\label{eq:koczkacl}
\end{equation}
with $z_i=\pm 1$ for all $i<m_B$. The expression is permutation
invariant; therefore it only depends on the number of $z_i=+1$
values. Let it be $k$. Then $\sum_{i>j}|z_i-z_j|=2k(m_B-k)$,
because $|z_i-z_j|=2$ when $z_i\neq z_j$, which occurs $k(m_B-k)$
times; otherwise it is zero. At the same time
$|\sum_{i=1}^{m_B}z_i|=|k-(m_B-k)|=|m_B-2k|$, therefore,
$\bar{\cal B}=2k(m_B-k)+\gamma |m_B-2k|$. This expression has the
same value at $k$ and at $m_B-k$, therefore it is enough to
consider $k \leq m_B/2$. It is easy to show that $\bar{\cal B}$
takes its maximum at $k_{max}$, which is the non-negative integer
nearest to $(m_B-\gamma)/2$, and its maximum value, that is, the
classical limit is $ \bar{\cal
B}_{cl}=(m_B^2+\gamma^2-4\Delta^2)/2$, where
$\Delta=|(m_B-\gamma)/2-k_{max}|$. For $\gamma>m_B$ this value is
$\gamma m_B$.

As we have shown, to get the maximum value for this family of Bell
inequalities we need $m_B$-dimensional vector spaces. With
$(m_B-1)$-dimensional spaces we may only get a smaller value, but
it is interesting to know how much smaller. In the following this
gap will be determined for particular number of $m_B$'s.

\subsection{Two-dimensional witnesses}\label{twodim}

We have determined numerically the maxima achievable in
$(m_B-1)$-dimensional spaces for small $m_B$ cases. The solutions
are nontrivial, they have different structures for different
ranges of $\gamma$. As three-dimensional spaces correspond to
qubits \cite{Brunner,VP08}, the Bell expression~(\ref{Bgamma})
defined above with $m_B=4$ and $m_A=7$ are especially interesting.
They have the smallest number of measurement settings for one of
the parties among correlation type Bell expressions (i.e.,
involving only joint correlations), whose maximum violation can
not be obtained with qubits, while the other participant has just
as few measurement settings as absolutely necessary. The maximum
value that can be achieved with three-dimensional vectors (i.e.\
with qubits when working with tensor products of Hilbert spaces)
as a function of $\gamma$ consists of three regions. For
$\gamma=0$ the four vectors $\vec b_i$ point toward the vertices
of a regular simplex. For finite $\gamma$ this simplex becomes
somewhat distorted, its shape will be a pyramid, whose base is a
regular triangle and whose apex is above the center of the base.
The maximum is given by such a solution up to $\gamma=1.4153$, a
value very near to $\sqrt{2}$, where a less regular shape takes
over. This is where the three-dimensional maximum value differs
the most from the global, the four-dimensional one, their ratio
here is 1.0107161. In the third region, valid for larger $\gamma$,
the four $\vec b_i$ vectors point toward the corners of the
square.

Remembering (according to remarks at the end of Sec.~\ref{rep})
that real qubits correspond to two-dimensional Euclidean spaces
and recalling the proof of Sec.~\ref{limits} it follows that
correlation Bell inequalities with $m_A,m_B\le 3$ do not require
complex Hilbert spaces for their maximal violation. On the other
hand, Bell inequality~(\ref{Bgamma}) with $m_B=3$ and $m_A=4$
and with $\gamma=1$ does require complex numbers to achieve
maximum violation (in the Euclidean space Bob's optimal vectors
form an orthogonal triad). Note, that this inequality is just the
same as the $3\times 4$ setting elegant Bell inequality introduced
in \cite{BG}. From Eq.~(\ref{Bmax}) it follows that the maximum
value is 6 in $\RR^3$, corresponding to complex qubits. On the
other hand, the maximum restricted to $\RR^2$, i.e., to real
qubits, can be obtained through semidefinite programming
\cite{VB96}. Expression (\ref{Bgamma}) to be maximized can be
brought to a form containing only quadratic variables subject to
quadratic constraints (actually we have 14 variables with 7
constraints). This non-convex problem can be solved via a series
of convex relaxations of increasing size \cite{Lasserre}. This
technique provides in each step a better upper bound to the global
optimum. We obtain the upper bounds of 6 in the first order and
5.8894 in the second order, which latter value coincides up to
numerical precision with the value which can be attained
numerically. We used the GloptiPol~3 package \cite{glopti} to
solve this global optimization problem. This approach thus gives
the gap between the maximum quantum value achievable in the
complex and real qubit spaces through an exact treatment (the
ratio being $6/5.8894 = 1.01878$). The gap takes its maximum at
$\gamma=1.3946$ with a ratio of $1.020 8047$. On the other hand,
using a heuristic method for the $m_B=5$ and $m_A=11$, and for the
$m_B=6$ and $m_A=16$ cases the maximum ratios are $1.006 2317$ (at
$\gamma=1.6396$), and $1.004 1964$ (at $\gamma=1.7642$),
respectively.

Now let us come back to Bell expression~(\ref{Bgamma}) with
$m_B=4$ and $m_A=7$. From our proof it does not follow that this
is actually the one with the minimum total number of measurement
settings whose maximum violation requires higher-dimensional
spaces than qubits. If we allow both participants to have more
than the minimum number of measurements, the sum may be decreased,
with $m_B=m_A=5$ it is just $10$. We have generated and checked
numerically all 44685 nonequivalent nontrivial inequalities with
$M_{ij}$ values restricted to $0$, $1$, and $-1$, and we have found
that qubits were enough to get the maximum value for each of them.
We then allowed $M_{ij}$ to be $0$, $1$, $-1$, $2$ and $-2$, while
confining ourselves to symmetric matrices. We found no case
requiring more than qubits for maximum violation even among these
7.66 million inequivalent cases. Although this still does not
prove that there is no $m_B=m_A=5$ correlation type Bell
inequality with this property, but it makes it likely. With
$m_A+m_B=11$ we found several examples even with allowing only
$0$, $1$ and $-1$ values for $M_{ij}$. With $m_B=4$ and $m_A=7$ we
found $11$ inequivalent cases (including the one discussed above
with $\gamma=1$), and with $m_B=5$ and $m_A=6$ our extensive
search gave 79 examples. Note that in these cases all of our
results are due to numerical search, since the complexity of this
particular problem was too large to be handled by the semidefinite
programming approach discussed previously. We are quite
confident, though, about the results obtained by heuristic numerical
computations.

The necessity of four-dimensional Euclidean space in achieving the
maximum value means that in the Hilbert space picture a pair of
qubits is not enough \cite{Brunner,VP08}. According to the
construct of Tsirelson \cite{Tsirelson} (see paragraph above
Lemma~\ref{lemmasharp}), any value one may achieve in
four-dimensional Euclidean spaces, one can also get with
measurement operators in four-dimensional complex Hilbert spaces.
We have calculated the maximum violation of the 79 $m_B=5$ and
$m_A=6$, the 11 $m_B=4$ and $m_A=7$ inequalities mentioned in the
previous paragraph, the $\gamma$-dependent $m_B=4$ and $m_A=7$
case for a few $\gamma$-values, and for the $m_B=4$ and $m_A=8$
example $X_4$ introduced in Ref.~\cite{VP08} with determining the
appropriate measurement operators via numerical optimization
according to Refs.~\cite{PV08a,PV08b}. In all cases complex
four-dimensional Hilbert spaces were required, smaller spaces were
never enough.

\section{Bounds on dimensions with infinite number of settings}\label{cont}

In this section we consider a bipartite Bell inequality with a
continuum infinite number of two-outcome measurement settings for
each party, which will be proved to serve as dimension witnesses
for arbitrary dimensions. Let the indices of the measurement
settings be $m$-dimensional unit vectors, that is, elements of
$S^{m-1}$, the surface of the unit sphere in $R^m$. Let the
Bell coefficients be proportional to the inner product of their
indices,
\begin{equation}
M(x,y)\equiv m\langle x,y\rangle \label{grot}
\end{equation}
where the factor of $m$ has been introduced for the sake of
convenience. By construction the coefficient matrix $M(x,y)$ is
positive semidefinite.  Eq.~(\ref{grot}) has been worked out by
Grothendieck \cite{Grothendieck} in context of functional
analysis. We use it in the following in matrix version formalized
in \cite{JP68}. The Bell expression with $n$-dimensional Euclidean
vectors will then be:
\begin{equation}
{\cal B}\equiv\int_{S^{m-1}}d\sigma(x)\int_{S^{m-1}}d\sigma(y) m\langle x,y\rangle \vec a(x)\cdot\vec
b(y), \label{Bgrot}
\end{equation}
where $\sigma$ denotes the normalized surface of the
$m$-dimensional unit sphere $S^{m-1}$, $\vec a(x)$, ($x\in
S^{m-1}$) and $\vec b(y)$, ($y\in S^{m-1}$) are the
$n$-dimensional unit vectors corresponding to the measurement
operators of Alice and Bob, respectively. Like in the previous
sections, we denote the scalar product of these vectors with a
dot. Grothendieck constructed this example to provide a lower
bound for his constant \cite{Finch}. Actually, he calculated
expression~(\ref{Bgrot}) for a uniform distribution of $x,y\in
S^{m-1}$ for $m\rightarrow\infty$ but to our knowledge the
optimality of the solution for any $m$ and $n$ has not been proved
yet. Below we give an optimal solution for any value of $m$ and
$n$, and show that by fixing $m\rightarrow\infty$ the maximum
value in $\RR^n$ is strictly increasing in $n$. As discussed in
Lemma~\ref{lemmasharp}, this will provide us a dimension witness
for any finite dimension. The maximum value of the Bell expression
achievable with $n$-dimensional vectors will then be:
\begin{equation}
{T^n}=\sup_{\vec b(y)\in S^{n-1}} m\int_{S^{m-1}}d\sigma(x)\left|\int_{S^{m-1}}d \sigma(y)\langle
x,y\rangle\vec b(y)\right|,
\end{equation}
where $\vec a(x)$ has been chosen optimally, that is, parallel to
\begin{equation}
\vec h(x)\equiv\int_{S^{m-1}}d\sigma(y)\langle x,y\rangle\vec b(y). \label{eq:vech}
\end{equation}

In the following an exact value will be given for the maximum
value of this Bell expression for any dimension $n$ in the
Euclidean space with respect to the classical case $\vec a, \vec
b=\pm1$ (i.e., setting $n=1$). Due to the lemma of Rietz
\cite{Rietz} this ratio can maximally be $\pi/2$ for a positive
semidefinite matrix $M$. Due to Grothendieck \cite{Grothendieck}
this ratio can be achieved for infinite $m,n$. This proves
incidentally that the solution in infinite $m,n$ is optimal.
However, next we can calculate exactly the optimal values $T^n$
for any $n$ and $m$ in contrast to the case in Sec.~\ref{bell}.

Due to the linearity of $\vec h(x)$, for each of its components
$h_i(x)$ there exist a $z_i\in S^{m-1}$ and a scalar $\lambda_i$
\cite{DavieReeds} such that
\begin{equation}
h_i(x)=\lambda_i\langle x,z_i\rangle
\end{equation}
The $h_i(x)$ is maximal at $z_i$, and its value there is $\lambda_i$:
\begin{equation}
\lambda_i=h_i(z_i)=\int_{S^{m-1}}d\sigma(y)\langle z_i,y\rangle b_i(y). \label{eq:lambdai}
\end{equation}
From these it follows that the maximum value of $\cal B$ in
$\RR^n$ may be written as:
\begin{equation}
{T^n}=\sup_{\vec b(y)\in S^{n-1}} m\int_{S^{m-1}}{d\sigma(x) \sqrt{\sum_{i=1}^n\lambda_i^2\langle
x,z_i\rangle^2}} \label{eq:calB}
\end{equation}

Let us introduce the following generalized spherical polar
coordinates for the components of unit vector $x$:
\begin{align}
\label{eq:pcoord}
x_1&=\cos\varphi_1\nonumber\\
x_i&=\cos\varphi_i\prod_{\mu=1}^{i-1}\sin\varphi_{\mu}\quad(1<i<m)\\
x_{m}&=\prod_{\mu=1}^{m-1}\sin\varphi_{\mu}\nonumber.
\end{align}
The integral of a function ${\cal F}(x)$ on the normalized unit sphere is:
\begin{align}
&\int_{S^{m-1}}d\sigma(x){\cal F}(x)=\int_0^{2\pi}d\varphi_{m-1}
\int_0^{\pi}d\varphi_{m-2}\sin\varphi_{m-2}\dots\nonumber\\&
\int_0^{\pi}d\varphi_{i}\sin^{m-i-1}\varphi_{i}\dots
\int_0^{\pi}d\varphi_{1}\sin^{m-2}\varphi_{1}{\cal F}{1\over{2s_0s_1\dots s_{m-2}}}, \label{eq:polint}
\end{align}
where $s_i\equiv\int_0^{\pi}\sin^i\varphi d\varphi= \sqrt\pi\Gamma((i+1)/2)/\Gamma((i+2)/2)$.

Let us consider the case of $n\le m$. Let us choose the basis such
that the last $n$ unit vectors $e_{m-n+1}$, $e_{m-n+2}$, \dots,
$e_{m}$ span an $n$-dimensional subspace which contains all $z_i$.
The value of the expression (\ref{eq:calB}) does not change if we
replace $x$ in the integrand with its projection $P_n x$ onto this
subspace. Let us define $\|P_n x\|= \sqrt{\langle P_n x,P_n
x\rangle}$ and introduce
\begin{equation}
x'\equiv {P_n x\over \|P_n x\|}, \label{eq:xprime}
\end{equation}
a unit vector in the $n$-dimensional subspace. Then
\begin{equation}
T^n=\sup_{\vec b(y)\in S^{n-1}} m\int_{S^{m-1}}{d\sigma(x) \sqrt{\sum_{i=1}^n\lambda_i^2\langle
x',z_i\rangle^2}}\cdot \|P_n x\| \label{eq:calB1}
\end{equation}
With our choice of the basis one can easily see from
definition (\ref{eq:pcoord}) of the generalized polar coordinates
that $\|P_n
x\|=\sin\varphi_{m-n}\sin\varphi_{m-n-1}\dots\sin\varphi_1$,
depending only on the first $m-n$ angles. At the same time, $x'$
is independent of these variables, therefore, the two factors in
the integrand of equation (\ref{eq:calB1}) may be integrated
separately. The integral of the first factor will be an integral
on $S^{n-1}$, while the value of the integral of the second
factor, as one can easily verify by substituting $\cal F$ with
$\|P_n x\|$ in equation (\ref{eq:polint}), is $s_{m-1}/s_{n-1}$.

We will determine $T^n$ by constructing an upper limit, and then
finding an explicit solution which saturates it. Let us consider
\begin{equation}
U_{kn}=\int_{S^{k-1}}{d\sigma(x)\sqrt{\sum_{i=1}^n\lambda_i^2\langle x,z_i\rangle^2}}.
\end{equation}
The function depends on $\vec b(y)$ (through $\lambda_i$ and
$z_i$), and $T^n= ms_{m-1}/s_{n-1}\sup_{\vec b(y)\in
S^{n-1}}U_{nn}$ if $n\le m$, and $T^n=m\sup_{\vec b(y)\in
S^{n-1}}U_{mn}$ otherwise. By using $\int f(x)dx\le\sqrt{\int
dx\int f^2(x)dx}$ we get:
\begin{equation}
U_{kn}\le\sqrt{\int_{S^{k-1}}d\sigma(x)\sum_{i=1}^n\lambda_i^2\langle x,z_i\rangle^2}=\sqrt{{1\over
k}\sum_{i=1}^n\lambda_i^2}.
\end{equation}
To get the last expression the integral of $\langle
x,z_i\rangle^2$ on the $k$-dimensional sphere has been carried
out, which can easily be done using appropriate polar coordinates.
For $\sum_{i=1}^n\lambda_i^2$ we get
\begin{equation}
\sum_{i=1}^n\lambda_i^2= \int_{S^{m-1}}d\sigma(y)\sum_{i=1}^n\lambda_i\langle z_i,y\rangle b_i(y),
\end{equation}
by substituting one $\lambda_i$ factor in each term according to
equation (\ref{eq:lambdai}) and by changing the order of the
summation and the integral. We may increase the integrand by
replacing the vector $\vec b(y)$ by the unit vector parallel to
$\vec h(y)$, that is the vector whose components are
$\lambda_i\langle z_i,y\rangle$. We may not necessarily be allowed
to choose $\vec b(y)$ this way, as $\lambda_i$ and $z_i$ are
determined by $\vec b(y)$, and this choice may be inconsistent.
However, we can overestimate the integral by making this
replacement, therefore:
\begin{equation}
\sum_{i=1}^n\lambda_i^2\le \int_{S^{m-1}}{d\sigma(y) \sqrt{\sum_{i=1}^n\lambda_i^2\langle
z_i,y\rangle^2}}.
\end{equation}
Comparing the right hand side of this inequality with equation
(\ref{eq:calB}) we can see that the maximum value it may take is
nothing else but $T^n$ divided by $m$. Therefore, it follows that
if $n\le m$,
\begin{equation}
T^n\le {ms_{m-1}\over s_{n-1}}\sqrt{{1\over n}{1\over m} T^n},
\label{eq:uplim0}
\end{equation}
that is,
\begin{equation}
T^n\le {m\over n}{s_{m-1}^2\over s_{n-1}^2}= {s_{m-1}\over s_{m}}{s_{n}\over s_{n-1}}.
\label{eq:uplim}
\end{equation}
For the last form we used the identity $ns_ns_{n-1}=2\pi$.

Let us choose $\vec b(x)$ to be the normalized projection of $x$
onto the $n$-dimensional subspace of the $m$-dimensional space
spanned by the last $n$ members of the basis introduced earlier,
that is
\begin{align}
b_i(x)&={\langle e_{m-n+i},x\rangle\over\sqrt{\sum_{j=m-n+1}^m\langle e_j,x\rangle^2}}
={\langle e_{m-n+i},x\rangle\over\|P_n x\|}\nonumber\\
&=\cos\varphi_{m-n+i}
\prod_{\mu=m-n+1}^{m-n+i-1}\sin\varphi_\mu.
\end{align}
For the last form we used the form of generalized polar
coordinates (\ref{eq:pcoord}), and that $\|P_n
x\|=\prod_{\mu=1}^{m-n}\sin\varphi_\mu$. To get $h_i(e_j)$ we have
to integrate the product of $\langle
e_j,x\rangle=\cos\varphi_j\prod_{\mu=1}^{j-1}\sin\varphi_\mu$ and
$b_i(x)$ on $S^{m-1}$ (see equation (\ref{eq:vech})). From
equation (\ref{eq:polint}) one can see that for $\varphi_j$ and
for $\varphi_{m-n+i}$, we have to integrate the product of the
cosine function and a power of the sine function if $j\neq m-n+i$. 
Therefore $h_i(e_j)=0$ for these values of $j$. From this it
follows that $z_i=e_{m-n+i}$, being $h_i(x)$ maximal for that
vector. Then we can explicitly calculate $\lambda_i$ from equation
(\ref{eq:lambdai}) by integrating $\langle e_{m-n+i},x\rangle
b_i(x)=\prod_{\mu=1}^{m-n}\sin\varphi_\mu
\prod_{\nu=m-n+1}^{m-n+i-1}\sin^2\varphi_\nu\cos^2_{m-n+i}$ on
$S^{m-1}$. Substituting this into equation (\ref{eq:polint}),
performing the integrations and simplifying the expression by
$\prod_{\tau=m-n+i+1}^{m-1}s_{m-\tau-1}$, we get
\begin{align}
\lambda_i&={\prod_{\mu=1}^{m-n}s_{m-\mu}\prod_{\nu=m-n+1}^{m-n+i-1}s_{m-\nu+1}(s_{n-i-1}-s_{n-i+1})
\over \prod_{\tau=1}^{m-n+i}s_{m-\tau-1}}\nonumber\\
&={s_{m-1}\over s_{n-1}}{s_{n}s_{n-1}\over s_{n-i+1}s_{n-i}}{s_{n-i-1}-s_{n-i+1}\over s_{n-i-1}}=
{s_{m-1}\over ns_{n-1}}.
\label{eq:lambdaicalc}
\end{align}
For the last equality the identities $s_n/s_{n-2}=(n-1)/n$ and
$ns_ns_{n-1}=2\pi$ have been used. Thus $\lambda_i$ is independent
of $i$, as it has to be due to symmetries. Then by bringing
$\lambda=\lambda_i$ in front of the integral in equation
(\ref{eq:calB}) and using $z_i=e_{m-n+i}$ we can see that the
integrand remaining is nothing else but $\|P_n x\|$, whose
integral is $s_{m-1}/s_{n-1}$, as we have seen earlier. The result
we get then for $T^n$ is equal to upper limit
(\ref{eq:uplim}).

For $n\ge m$, it is easy to verify that the upper limit is $1$
(instead of equation (\ref{eq:uplim0}) we get $T^n\le\sqrt{T^n}$),
which is the well-known quantum limit for this inequality. It can
be reached with $m$-dimensional vector space, with $\vec b(x)=x$.
As the limit remains the same for $n>m$, there is no need for the
extra dimensions. Although the Bell inequality involves an
infinite number of measurement settings for each party, a finite
dimensional space is enough to reach the quantum limit.

Now we set $m\rightarrow\infty$ in (\ref{eq:uplim}) and calculate
$T^n$ for different $n$ values. In this case $T^n = s_n/s_{n-1}$.
By choosing $n=1$ we obtain the known classical limit $2/\pi$.
Thus the maximum quantum violation (quantum limit per classical
limit) is $ (\pi/2) s_n/s_{n-1}$. For $n=2$ the ratio is
$\pi^2/8\approx 1.2337$, corresponding to measurement on pairs of
real qubits, while with $n=3$, corresponding to complex qubits, it
is $4/3$. We have to go up to $n=5$ to get a maximum
violation of $64/45\approx 1.4222$, larger than the value of $\sqrt 2$
one can achieve with the CHSH inequality. For $n\rightarrow\infty$
the maximum violation is the well-known $\pi/2$, which is
$3\pi/8\approx 1.1781$ times larger than that we can achieve with
qubits.

Most importantly the ratio $s_n/s_{n-1}$ is a strictly increasing
function of $n$. Notably for $n$ even it is
$(\pi/4)\prod_{i=1}^{n/2-1}{(2i+1)^2/((2i+1)^2-1)}$ and for $n$
odd it is equal to
$(2/\pi)\prod_{i=1}^{(n-1)/2}{(2i)^2/((2i)^2-1)}$. Clearly both
are strongly monotone functions of $n$. This entails $T^n<T^{n+1}$
by $m\rightarrow\infty$ for all $n$ and owing to
Lemma~\ref{lemmasharp} proves the existence of dimension witnesses
for any finite dimension.

\section{Conclusion}\label{conc}

In the present work we focused our attention on joint quantum
correlations arising from local measurements on bipartite systems.
Tsirelson has established a connection between these joint
correlations in Hilbert space and standard inner products of unit
vectors in Euclidean space. Based on this result we give a proof
that constructing a correlation Bell expression for which
$T^n<T^{n+1}$ in the vectorial picture (where $T^n$ denotes the
maximum value achievable in the $n$-dimensional Euclidean space)
implies joint correlations which cannot be reproduced in a
$d$-dimensional Hilbert space, where
$d=\lfloor\frac{n+1}{2}\rfloor$. This defines a $d$-dimensional
witness. For this sake, we discuss two particular families of Bell
inequalities. The one in Sec.~\ref{finite} involves a finite number of
measurement settings $m_B$ and $m_A= m_B(m_B-1)/2 + 1$, whereas
the other one in Sec.~\ref{cont} involves continuously many settings
on both sides. Though in the former case we cannot give the
difference $T^{n+1}-T^n$ explicitly, except for small values of
$m_B$, in the latter case this difference can be analytically
calculated for all $n$. This conclusively proves the existence of
dimension witnesses for arbitrary dimension $d$ in a bipartite
quantum system. Recently, we learned that Bri\"{e}t et al. arrived 
in Ref.~\cite{Briet} at similar conclusions. Besides, we discuss the 
minimum number of measurement settings arising in correlation Bell 
expressions in order to generate dimension witnesses. In this respect 
we prove that the numbers of settings $m_B=4$ and $m_A=6$ (or less 
settings) are not sufficient to generate a two-dimensional witness. 
In contrast, examples are given for a two-dimensional witness for 
settings $m_B=4$ and $m_A=7$ and for settings $m_B=5$ and $m_A=6$. 
It remains an open question whether for the pair of settings $m_B=5$
and $m_A=5$ there exists a two-dimensional dimension witness or not.

Due to Tsirelson's work \cite{Tsirelson} for a given finite number
of binary measurement settings, it is always possible to generate
all the joint bipartite quantum correlations of binary outcomes in
a finite-dimensional Hilbert space. However, in generic Bell
expressions involving local marginal terms as well, this may not
be true. With respect to it, Navascu\'{e}s et al. \cite{NPA08}
asked recently whether there exist scenarios with finite number of
measurement settings for both parties, for which all quantum
correlations can be attained by measuring an infinite-dimensional
entangled system. We leave this interesting problem as a challenge
for future investigations.

\acknowledgments T.V. was supported by a J\'anos Bolyai Grant
of the Hungarian Academy of Sciences.

\end{document}